\documentclass[journal, twoside]{IEEEtran}

\usepackage[english]{babel}
\usepackage{fancyhdr}
\usepackage{lipsum}
\usepackage[utf8]{inputenc}
\usepackage[T1]{fontenc}
\usepackage{graphicx}
\usepackage{nccmath}
\usepackage[table]{xcolor}
\usepackage{cite}
\usepackage{acro}
\usepackage{scalerel}
\usepackage{accents}
\usepackage{tikz}
\usetikzlibrary{svg.path}
\usepackage{amsmath}
\usepackage{amssymb}
\usepackage{amsthm}
\usepackage{mathtools}
\usepackage{algorithmicx}
\usepackage{array}
\usepackage[caption=false,font=footnotesize]{subfig}
\usepackage{url}
\usepackage{hyperref}
\usepackage[english, capitalise]{cleveref}

\graphicspath{{../figure/}}

\addto\captionsbrazilian{\renewcommand{\figurename}{Fig.}}
\addto\captionsenglish{\renewcommand{\figurename}{Fig.}}

\definecolor{orcidlogocol}{HTML}{A6CE39}
\tikzset{
  orcidlogo/.pic={
    \fill[orcidlogocol] svg{M256,128c0,70.7-57.3,128-128,128C57.3,256,0,198.7,0,128C0,57.3,57.3,0,128,0C198.7,0,256,57.3,256,128z};
    \fill[white] svg{M86.3,186.2H70.9V79.1h15.4v48.4V186.2z}
                 svg{M108.9,79.1h41.6c39.6,0,57,28.3,57,53.6c0,27.5-21.5,53.6-56.8,53.6h-41.8V79.1z M124.3,172.4h24.5c34.9,0,42.9-26.5,42.9-39.7c0-21.5-13.7-39.7-43.7-39.7h-23.7V172.4z}
                 svg{M88.7,56.8c0,5.5-4.5,10.1-10.1,10.1c-5.6,0-10.1-4.6-10.1-10.1c0-5.6,4.5-10.1,10.1-10.1C84.2,46.7,88.7,51.3,88.7,56.8z};
  }
}

\newcommand\orcidicon[1]{\href{https://orcid.org/#1}{\mbox{\scalerel*{
\begin{tikzpicture}[yscale=-1,transform shape]
\pic{orcidlogo};
\end{tikzpicture}
}{|}}}}

\DeclareAcronym{acm}{
  short = ACM ,
  long  = Association for Computing Machinery ,
  sort  = A
}

\DeclareAcronym{ufrj}{
  short = {UFRJ},
  long  = {Universidade Federal do Rio de Janeiro},
  sort  = U
}

\DeclareAcronym{uff}{
  short = {UFF},
  long  = {Universidade Federal Fluminense},
  sort  = U
}

\DeclareAcronym{uerj}{
  short = {UERJ},
  long  = {Universidade do Estado do Rio de Janeiro},
  sort  = U
}

\DeclareAcronym{ence}{
  short = {ENCE},
  long  = {Escola Nacional de Ciências Estatísticas},
  sort  = E
}

\DeclareAcronym{ibge}{
  short = {IBGE},
  long  = {Instituto Brasileiro de Geografia e Estatística},
  sort  = I
}

\definecolor{Pos01}{HTML}{082C59}
\definecolor{Pos02}{HTML}{3777BC}
\definecolor{Pos03}{HTML}{76B4E8}
\definecolor{Pos04}{HTML}{B9DEF7}
\definecolor{Pos05}{HTML}{F9F8F9}
\definecolor{Pos06}{HTML}{FFCABD}
\definecolor{Pos07}{HTML}{FF8D7B}
\definecolor{Pos08}{HTML}{D14136}
\definecolor{Pos09}{HTML}{610102}

\definecolor{hr}{HTML}{C721DD}
\definecolor{halpha}{HTML}{D14A00}
\definecolor{hbeta}{HTML}{008C00}
\definecolor{hgamma}{HTML}{007FB1}
\definecolor{hdelta}{HTML}{D1AC00}
\definecolor{hphi}{HTML}{870036}
\definecolor{hBeta}{HTML}{2EFF71}
\definecolor{hGamma}{HTML}{00D0D9}
\definecolor{hPhi}{HTML}{F1004F}

\newcommand\crule[3][black]{\textcolor{#1}{\rule{#2}{#3}}}

\crefname{heuristic}{Heuristic}{Heuristics}
\creflabelformat{heuristic}{#2\textup{#1}#3}

\crefname{step}{Step}{Steps}
\creflabelformat{step}{#2\textup{#1}#3}

\begin{document}


\title{Spectral Heuristics Applied to Vertex Reliability}


\author{Carla Silva Oliveira \orcidicon{0000-0001-6684-8811}\,, Fausto Marques Pinheiro Junior \orcidicon{0000-0001-9727-8532}\,, and José André de Moura Brito \orcidicon{0000-0002-2825-0058}
\thanks{Carla S. Oliveira is with Escola Nacional de Ciências Estatísticas, Instituto Brasileiro de Ciências Estatísticas, Brasil, e-mail: carla.oliveira@ibge.gov.br.}
\thanks{Fausto M. Pinheiro Jr. is with Programa de Pós-Graduação em Matemática Aplicada, Universidade Estadual de Campinas, Brasil, e-mail: fausto.mpj@protonmail.com}
\thanks{José A. M. Brito is with Escola Nacional de Ciências Estatísticas, Instituto Brasileiro de Ciências Estatísticas, Brasil, e-mail: jose.m.brito@ibge.gov.br.}

\thanks{Draft Version. Last revision: Month 11, 2022}}



\markboth{}
{}  




\maketitle

\begin{abstract}
    The operability of a network concerns its ability to remain operational, despite possible failures in its links or equipment. One may model the network through a graph to evaluate and increase this operability. Its vertices and edges correspond to the users' equipment and their connections, respectively. In this article, the problem addressed is identifying the topological change in the graph that leads to a greater increase in the operability of the associated network, considering the case in which failure occurs in the network equipment only. More specifically, we propose two spectral heuristics to improve the vertex reliability in graphs through a single edge insertion. The performance these heuristics and others that are usually found in the literature are evaluated by computational experiments with $22000$ graphs of orders $10$ up to $20$, generated using the Models Erd\H{o}s-Rényi, Barabási-Albert, and Watts-Strogatz. From the experiments, it can be observed through analysis and application of statistical test, that one of the spectral heuristics presented a superior performance in relation to the others.
\end{abstract}

\begin{IEEEkeywords}
    Graph, Vertex Reliability, Spectral Heuristic.
\end{IEEEkeywords}

\IEEEpeerreviewmaketitle



\section{Introduction}\label{Intro}

\IEEEPARstart{T}{he} quality of mass telecommunication services is a key point for companies to prosper as technology dominates a competitive globalized market\cite{sp_2019}. Such services are well-known examples of network structures composed of equipment (e.g., routers) and links (e.g., cables) that can connect millions of users.

The capacity of a network to remain operational despite possible (random) failures in some of its equipment or links is called \textit{operability}. Therefore, it is important to consider the issue of operability to design networks that provide high quality of services. Similar to other problems associated with networks, a new approach was developed to solve the problem of maximizing the operability of a system by modelling its structure through a graph.

Graphs are mathematical objects that allow modelling a wide range of phenomena to identify properties due to their structure\cite{newman_2010}. One of such properties is called \textit{reliability} -- the probability of a (connected) graph remaining connected after the (random) removal of an arbitrary number of vertices or edges. In particular, if only the vertices can be removed, then this property is called \textit{vertex reliability}. Analogously, we have the \textit{edge reliability}.  Thus, the reability of the graph that models the network enables the investigation of the operability of the network. 

Theoretically, the graph that models the network can pinpoint which changes to its topology lead to maximal reliability, even when given a limited number of changes. Consequently, these changes can be mapped to adjustments in the network structure that maximize the operability under any desired constraints. However, calculating the reliability of a graph belongs to the class of NP-Hard problems, which makes its exact solution unfeasible in practice for most real networks\cite{sutner_1991}. Therefore, determining which topological change causes the greatest increase in the reliability of a graph is usually made through heuristics based on its topological properties.

This article specifically addresses the problem of identifying the edge insertion on a graph that causes the greatest increase in its vertex reliability. To achieve this goal, we develop and analyse two spectral heuristics, namely: \textbf{(a)} the heuristic of the greatest increase in algebraic connectivity ($\alpha$ heuristic); and \textbf{(b)} the heuristic of the greatest absolute difference between the components of the Fiedler vector ($\varphi$ heuristic). Both are not entirely original and were previously presented as a single heuristic in the context of edge reliability\cite{barreto_2017}. However, we propose a critique of this previous heuristic and the investigation of its algorithm and theoretical motivation. Thus, in addition to changing the context, we seek to split this previous heuristic and argue that an unexplored aspect in its algorithm makes it more appropriate than it should be, given the theoretical motivation that underpinned it. In particular, $\varphi$ outperformed $\alpha$ and other commonly used heuristics from the literature in a set of computational experiments.

This paper is organized as follows: \cref{Prelim} presents the main concepts of Graph Theory and Reliability employed in the work; \cref{Revisao} presents a review of the literature focusing on the contextualization of the research problem; \cref{Metodo} presents the descriptions and algorithms of the proposed heuristics; \cref{EC} presents the specification, mode of comparison and analysis of the computational experiments; and, finally, \cref{Conclusao} presents the conclusions and suggestions for future work.



\section{Preliminary Concepts}\label{Prelim}


\subsection{Graph Theory}\label{Prelim-grafos}

Consider a \textit{simple graph} $G=(V(G),E(G))$, such that $V(G) = \{v_{1}, \ldots, v_{n}\}$ is a finite set whose elements are called \textit{vertices}, and $E(G) \subseteq \{ \{ v_{i}, v_{j} \} : v_{i}, v_{j} \in V(G),\ i \neq j \}$ is a finite set whose elements are $2$-elements subsets of $V(G)$ and are called \textit{edges}, such that $\vert V(G) \vert = n$ and $\vert E(G) \vert = m$. If $V(H) \subseteq V(G)$ and $E(H) \subseteq E(G)$, then $H = (V(H), E(H))$ is a \textit{subgraph} of $G$ and $G$ is a \textit{supergraph} of $H$, $H \subseteq G$. In particular, if $E(H) = \{ \{v_{i}, v_{j}\} : v_{i}, v_{j} \in V(H), \{v_{i}, v_{j}\} \in E(G) \}$, then $H$ is an \textit{induced subgraph} of $G$. For non-adjacent vertices $v_{i}$ and $v_{j}$, $Y_{ij} = G + \{ v_{i}, v_{j} \}$ is a supergraph of $G$ obtained by the insertion of an edge between $v_{i}$ and $v_{j}$. The degree of a vertex $v_{i} \in V$, $d(v_{i})$, is the number of edges that are incident to it. If $d (v_{i}) = 1$, then $v_{i}$ is called a \textit{pendant vertex}.

A path is a graph $P$ such that $V(P) = \{v_{0},v_{1},  \ldots, v_{n}\}$ and $E(P) = \{ \{v_{0}, v_{1} \}, \{v_{1}, v_{2} \}, \ldots, \{v_{n-1}, v_{n} \} \}.$ In particular, two paths $P$ and $Q$ are \textit{vertex- or edge-independent} if $V(P) \cap V(Q) = \emptyset$ or $E(P) \cap E(Q) = \emptyset$, respectively. The \textit{length} of a path is its number of edges. The \textit{distance} between $v_{i}$ and $v_{j}$, $d_{G}(v_{i}, v_{j})$, is the number of edges in a shortest path between them, if such path exists. Otherwise, $d_{G}(v_{i}, v_{j}) = \infty$. The diameter of $G,$ $\text{diam}(G),$ is the maximum distance between all pair of vertices in $G$.  If $\vert V \vert = 1$ or if there is at least one path for every pair of vertices in $V$, then $G$ is connected. Otherwise, $G$ is disconnected.


The \textit{vertex connectivity} of $G$, $\kappa(G)$, is the smallest number of vertex-independent paths between all pairs of vertices in $V$. If $\kappa(G) = k$, then $G$ is $k$-connected. A $r$-\textit{vertex cut set} of $G$ is a set of $r$ vertices whose removal makes $G$ disconnected. If there is no $k$-vertex cut set for $0 < k < r$, then the $r$-vertex cut set is the \textit{minimum vertex cut set}.

The \textit{betweenness centrality} of a vertex $v_{i} \in V$, $g(v_i)$, is the ratio between the number of shortest paths between all pairs of vertices other than $v_{i}$ that contains $v_{i}$ and the total number of shortest paths between all pairs of distinct vertices.

The \textit{Laplacian matrix of G}, $L(G)=[l_{ij}]$, defined as follows:

    \begin{align*}
        l_{ij} =   \begin{cases}
                        d(v_{i}), & \text{if } i = j \\
                        -1, & \text{if } \{ v_{i}, v_{j}\} \in E(G) \\
                        0, & \text{otherwise}
                    \end{cases} \text{.}
    \end{align*}

The second smallest eigenvalue of $L(G)$, $\alpha(G)$, is called \textit{algebraic connectivity} of $G$. An eigenvector $\mathbf{\nu_{2}}(G)$ of $L(G)$ associated to $\alpha(G)$ is a \textit{Fiedler vector}. The \textit{Fiedler distance} between two vertices $v_{i}$ and $v_{j}$ of $G$ is the absolute difference between the $i$-th and $j$-th coordinate of the Fiedler vector, $d_{ \mathbf{ \nu_{2} } (G) } (v_{i}, v_{j}) = \vert (\mathbf{ \nu_{2} } (G))_{i} - (\mathbf{ \nu_{2} } (G))_{j} \vert$.


\subsection{Reliability}\label{Prelim-confiabilidade}

Let $G$ be a connected graph such that each vertex or edge is subject to removal with independent probability $1 - p$, $p \in [0,1]$. The reliability of $G$ is a polynomial function that maps $(G,p)$ to the probability that $G$ remains connected after the removal of a random number of its vertices or edges. Since Moore \& Shannon, the study of the reliability of a graph has focussed on the edge reliability, $R_{E}(G, p)$, which assumes reliable vertices and unreliable edges\cite{moore_1956}. Thus, a minor part of the literature turned to the study of the vertex reliability, $R_{N}(G, p)$, which assumes reliable edges and unreliable vertices\cite{goldschmidt_1994}.

\begin{figure}[!ht]
    \centering
    \includegraphics[width=2.0in]{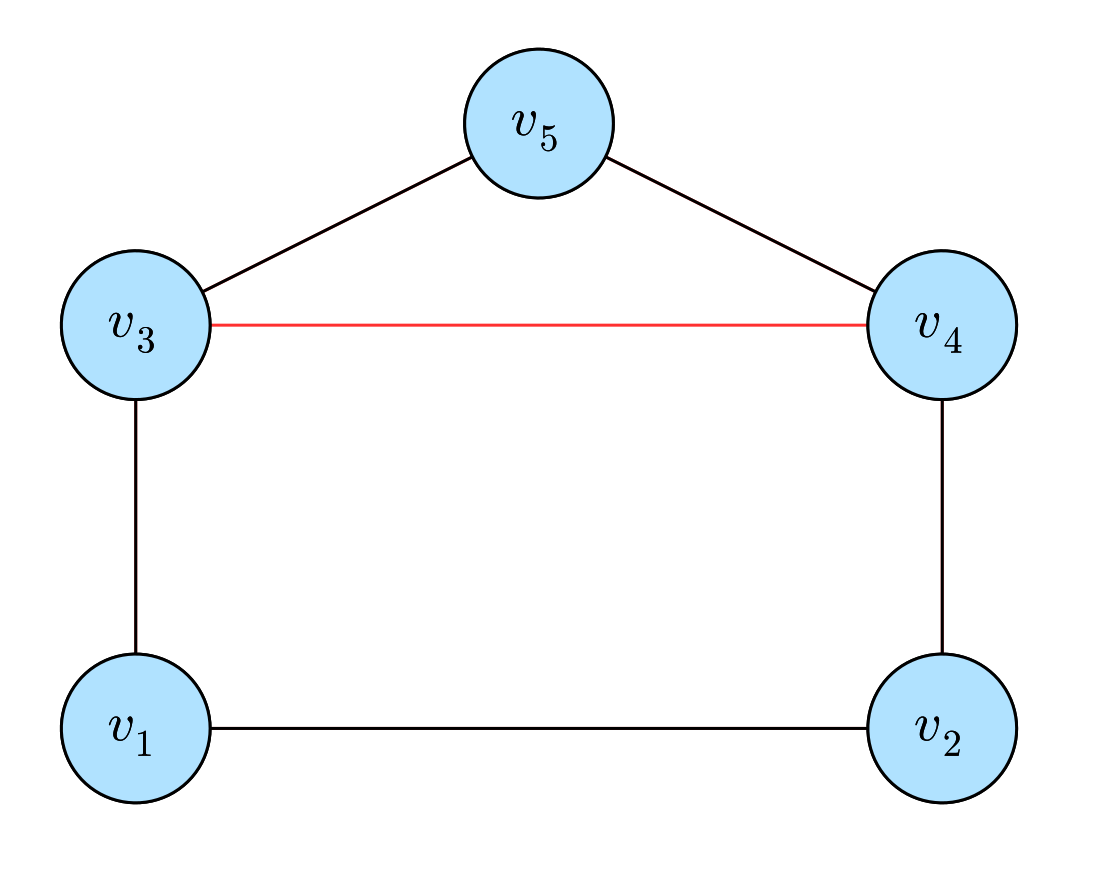}
    \caption{Graph $C_{5}$, supergraph $Y_{34}$, $R_{N}(G, 0.9) = 0.95949$ and $R_{N}(Y_{34}, 0.9) = 0.97488$.}
    \label{fig-casa}
\end{figure}

In both cases, determining the reliability of $G$ belongs to the class of NP-Complete problems; because the problem of counting the number of induced connected subgraphs of a given graph is NP-Complete \cite{sutner_1991}. Specifically, the vertex reliability polynomial of $G$ is given by


\begin{equation}\label{reliabilityfunction}
    R_{N}(G, p) = \sum^{n}_{r = 1} S_{r}(G) p^{r}(1 - p)^{n - r}  \text{ ,}
\end{equation}

\noindent with

\begin{equation}\label{binomialsum}
    S_{r}(G) + C_{n-r}(G) = \binom{n}{r}  \text{ ,}
\end{equation}

\noindent where $S_{r}(G)$ is the number of connected induced subgraphs of $G$ with exactly $r$ vertices, and $C_{n-r}(G)$ is the number of $(n-r)$-vertex cut sets of $G$\cite{liu_2000}. The \cref{fig-casa} shows the graph $C_{5}$, one of its supergraphs obtained by inserting the edge $\{v_{3},v_{4}\}$, $Y_{34}$, and the vertex reliability of $C_5$ and $Y_{34}$ given $p=0.9$.

Suppose $G \in \Omega(n,m)$, the set of all connected graphs with $n$ vertices and $m$ edges. Then, $G$ is: \textbf{(a)} \textit{uniformly best} if, and only if, $R_{N}(G,p) \geq R_{N}(H,p), \forall p \in [0,1], \forall H \in \Omega(n,m)$; \textbf{(b)} \textit{locally best} if, and only if, $\exists p \in [0,1] : R_{N}(G,p) \geq R_{N}(H,p), \forall H \in \Omega(n,m)$; \textbf{(c)} $\kappa$\textit{-optimal} if, and only if, $\kappa(G) \geq \kappa(H), \forall H \in \Omega(n,m)$; and \textbf{(d)} $S_{r}$\textit{-optimal}, for $1 \leq r \leq n$, if and only if, $S_{r}(G) \geq S_{r}(H), \forall H \in \Omega(n,m)$\cite{goldschmidt_1994}. Additionally, if $G$ is uniformly best, then it is $\kappa$-optimal; and if $G$ is $S_{r}$-optimal for all $r \in \{1,\ldots, n\}$, then it is uniformly best\cite{liu_2000}.

However, a uniformly best graph does not always exist. There are at least two locally best graphs in those cases, one of which is necessarily $kappa$-optimal\cite{goldschmidt_1994}. It is conjectured that a graph is at least locally best if it is $S_{r}$-optimal for a sufficient unknown number of $r \in \{1,\ldots,n\}$\cite{goldschmidt_1994}. Trivially, every $G \in \Omega(n,m)$ is $S_{1}$-optimal, $S_{2}-optimal$, and $S_{n}$-optimal. Moreover, by \cref{binomialsum} and Menger's Theorem, a $kappa$-optimal $G$ such that $\kappa(G) = k$ is $S_{n-k}$-optimal\cite{diestel_2017}. It is reasonable to suggest that if $H \in \Omega(n,m)$ is $S_{r}$-optimal in more values of $r$ than a $\kappa$-optimal graph $G$, then $H$ is a locally best graph.



\section{Literature Review}\label{Revisao}

Shier identifies two main areas of research related to reliability in graphs\cite{shier_1991}. The first one is \emph{analysis}, which involves calculating the reliability of a graph, analysing its polynomial function, and finding bounds for the values it can assume. The second one is \emph{design}, which involves constructing or changing a graph to achieve high reliability, usually through topological optimization and heuristics for insertion or rearrangement. This separation is not rigid despite the distinction, and many works deal with analysis and design. In any case, Colbourn emphasizes that both are primarily concerned with connectivity and its limitations due to topology.\cite{colbourn_1991}.

From the analysis perspective, part of the literature turned to the simulation of states of a graph to estimate its main topological properties\cite{shpungin_2012}. The primary tool within this subarea is the Monte Carlo simulation, for which several specific techniques have been developed\cite{lomonosov_1994, gertsbakh_2012}. Although these simulation techniques can produce satisfactory approximates to various reliability-related problems, the accuracy limitation inherent to simulations makes them an unfit tool to compare the effect of a single edge insertion on the reliability of an arbitrary graph\cite{fishman_2013}. The difference in reliability between single edge insertion may be smaller than what the simulations can identify.

From the design perspective, part of the literature tries to mitigate the difficulty of reliability by restricting its research to graphs with specific structures (e.g., almost regular complete multipartite graphs), using specific types of reliability (e.g., $s$-$t$ reliability), or limiting the types and quantities of allowed changes to a graph (e.g., rearranging at most $1$ edge). In effect, design is well-placed to develop heuristics for constructing or changing graphs.

These heuristics produce solutions that don't guarantee optimality in arbitrary graphs but are sufficiently good under the assumed constraints. That is, it is possible to stipulate the criteria of what constitutes a satisfactory solution, such as always being above average in comparison to arbitrary feasible solutions or by the proximity of the optimal in cases where it is possible to compare them with the \textit{optimum}.

Although there are different heuristics for distinct problems related to connectivity in general, most of them use similar algorithms based on the topological properties of graphs\cite{beygelzimer_2005, sydney_2013}. In those cases, the difference usually involves adapting which graph property is considered (e.g., vertex connectivity instead of edge connectivity) and weighting the quality of the solutions by their computational cost.

However, there are some peculiarities about vertex reliability that add difficulty to the investigation when compared to edge reliability. Notably among these: \textbf{(a)} $R_{N}(G,p)$ is non-monotonic while $R_{E}(G,p)$ is monotonic; \textbf{(b)} vertex connectivity and the enumeration of connected induced subgraphs are used instead of edge connectivity and the enumeration of spanning trees; and \textbf{(c)} most reduction techniques in the literature assume perfectly reliable nodes, such as deletion–contraction by the factoring theorem\cite{carlier_1991, torrieri_1994, lin_2007, brown_2018}. Thus, it is necessary to have different heuristics or ensure that adaptations take into account these peculiarities and test their performance in the context of vertex reliability.

The article addresses the problem of deciding which edge insertion leads to the greatest increase in vertex reliability on an arbitrary connected graph without pendant vertices. More specifically, we adapted and evaluated for vertex reliability six heuristics in the literature on edge reliability and other robustness measures. Then, based on the performance results and theoretical considerations, we focus our attention on the proposed $\alpha$ and $\varphi$ heuristics and how their behaviours suggest renewed approaches to spectral heuristics and reliability\cite{barreto_2017}.



\section{Method}\label{Metodo}

The heuristics evaluated in this work are divided between \textit{basic} heuristics, named by lowercase letters, and \textit{derivative} heuristics, named by uppercase letters. In all cases, a heuristic applied to a graph produces a solution in the form of single edge insertions to increase the vertex reliability of this graph.

The basic heuristics don't have an explicit tiebreak criterion with regard to the multiplicity of single edge insertions produced as a solution for a graph. That is, a basic heuristic applied to $G$ can produce more than one single edge insertion, and such insertions are considered equivalent to each other. In turn, each derivative heuristic is a refinement of a basic heuristic by adding a tiebreak criterion to guarantee the unicity of the output.

The $\alpha$ and $\varphi$ heuristics we propose in this work are described by their: \textbf{(a)} theoretical motivation; and \textbf{(b)} algorithm, that is, the steps and calculations prescribed by the heuristic. All the other heuristics are briefly addressed in \cref{Metodo-demais}.


\subsection{Proposed Heuristics}\label{Metodo-heuristicas}


\subsubsection{$\alpha$ heuristic}\label{Metodo-alpha}

The theoretical motivation for the $\alpha$ heuristic is the maximization of the algebraic connectivity of $G$\cite{barreto_2017}. Consider $\mathcal{Y} = \{ Y_{ij} = G + \{i,j\} : G \in \Omega(n,m), v_{i} \not\sim v_{j} \}$, the set of all supergraphs of $G$ generated by a single edge insertion. Since $\mathcal{Y} \subset \Omega(n, m+1)$, it is necessary that if $Y \in \mathcal{Y}$ is locally best in $\Omega(n, m+1)$, then it must have the maximum vertex connectivity of all graphs in $\mathcal{Y}$. Thus, the optimal increase in vertex connectivity is a necessary but insufficient criterion for a supergraph to be locally best. Moreover, given that $\alpha(G) \leq \kappa(G)$ for $G$ with $n>1$ vertices, then the algebraic connectivity is positively correlated with the vertex connectivity\cite{fiedler_1973}. Therefore, it is reasonable to ponder if a heuristic that increases the algebraic connectivity might also consistently increase the vertex reliability.

The algorithm for the $\alpha$ heuristic involves calculating $\alpha(Y_{ij})$ for every possible edge insertion $\{v_{i}, v_{j}\}$ of $G$, sorting these values, and choosing the edge insertions associated with the maximum values. Note that if $G$ is a connected graph with $n$ vertices, it necessarily has at least $(n - 1)$ edges, and at most $\binom{n}{2}$ edges. Thus, the minimum number of edge insertions is equal to $0$, and the maximum number is $\binom{n}{2} - n + 1$. The calculation of the eigenvalues of a single $n \times n$ matrix belongs to $O(n^{2})$\cite{pan_1998}. Therefore, calculating the algebraic connectivity for every supergraph of a sparse graph is computationally expensive.


\subsubsection{$\varphi$ Heuristic}\label{Metodo-phi}

Given that the $\alpha$ heuristic could be too costly if it calculated the algebraic connectivity for every supergraph of $G$, the literature suggests the use of a heuristic that maximizes the algebraic connectivity instead of the exact method described in \cref{Metodo-alpha}\cite{barreto_2017}. The Fiedler distance of an edge insertion is a theoretically sound and reasonable proxy for increases in algebraic connectivity, due to the generally positive correlation between them in single edge insertions and $\alpha(G) \leq \alpha(Y_{ij}) \leq ( d_{ \mathbf{ \nu_{2} } (G) } (v_{i}, v_{j}) )^{2} + \alpha(G)$\cite{godsil_2001}. It is in this context that the $\varphi$ heuristic first appeared\cite{wang_2008}.

It must be noted that too many steps were made while searching for a heuristic for vertex reliability. Initially, a good candidate was identified -- vertex connectivity. However, it is computationally expensive. The algebraic connectivity is a good proxy for it, though, but also too computationally expensive. Finally, Fiedler distance is a good proxy for the algebraic connectivity, so it is possible to wonder if a proxy (Fiedler distance) for a proxy (algebraic connectiviy) of a proxy (vertex connectivity) is, in fact, a good heuristic for vertex reliability. Surprisingly, it is.

The correlation between Fiedler distance in single edge insertions and increase in algebraic connectivity in an arbitrary graph depends on its topology. Although the correlation is generally positive, it shows considerable variability when different classes of random graphs with distinct topological properties are considered. In the worst-case scenario described in the literature, the correlation drops to $0$ in highly regular $D$-lattices\cite{wang_2008}. If $\varphi$ is a good heuristic due to the fact that it is an excellent proxy for algebraic connectivity, then in those cases it would perform poorly.

We propose a theoretical motivation for $\varphi$ heuristic based on Fiedler's Theorem, by which the number of connected induced subgraphs of $G$ increases as the maximum Fiedler distance between its pairs of vertices decreases\cite{fiedler_1973}. In \cref{Prelim-confiabilidade}, we postulated that a graph that is $S_{r}$-optimal in more values of $r$ than a $\kappa$-optimal graph is locally best. Assuming that this is true, having the greatest increase of connected induced subgraphs is a sufficient but unnecessary criterion for a supergraph to be locally best. Assuming that this is true, having the greatest increase of connected induced subgraphs is a sufficient but unnecessary criterion for a supergraph to be locally best.



In addition, Fiedler distance has two important properties: \textbf{(a)} if $\mathbf{ \nu_{2} } (G)$ is a unit Fiedler vector, then $d_{ \mathbf{ \nu_{2} } (G) } (v_{i}, v_{j}) \leq \sqrt{2}$; and \textbf{(b)} $d_{ \mathbf{ \nu_{2} } (Y_{ij}) } (v_{i}, v_{j}) \leq d_{ \mathbf{ \nu_{2} } (G) } (v_{i}, v_{j}) $. The first property ensures that Fiedler distance is limited; it is bounded between $0$ (lower bound) and $\sqrt{2}$ (upper bound) for an arbitrary $G$ of order $n$. The second ensures that the Fiedler distance between two non-adjacent vertices is always greater than or equal to the Fiedler distance between these same vertices on any supergraph that contain an edge incident to these two vertices.


The algorithm for the $\varphi$ heuristic involves calculating the eigenvalues and eigenvectors of $L(G)$, selecting the Fiedler vectors associated with $\alpha(G)$, and then finding the absolute difference between every pair of non-adjacent vertices. Algorithms for calculating the eigenvalues and eigenvectors of a square matrix are $O(n^{3})$\cite{pan_1998}. If the geometric multiplicity of $\alpha(G)$ is greater than $1$, then each pair of vertices has its Fiedler distance evaluated in all Fiedler vectors.

As a derivative heuristic of $\varphi$, we also propose the heuristic $\Phi$ which presents the following tiebreak criterion: if there are multiple single edge insertions with the same Fiedler distance, then the chosen insertion will be the one that maximizes the algebraic connectivity of its supergraph. 


\subsection{Other Heuristics}\label{Metodo-demais}


\subsubsection{$\beta$ Heuristic}\label{Metodo-beta}

The $\beta$ heuristic inserts a single edge between non-adjacent vertices with minimum betweenness centrality. Its theoretical motivation is to reduce the mean betweenness centrality of the vertices of $G$\cite{holme_2002, jiang_2011, ji_2016}. Given that a relatively high betweenness centrality of a vertex $v_{i}$ is positively correlated with the probability that $v_{i}$ belongs to a minimal vertex cut set, it is reasonable to ponder whether a heuristic that minimizes the mean betweenness centrality may also increase the vertex reliability. As a derivative heuristic of $\beta$, we postulate a $B$ heuristic. This heuristic is ideal, as it has no operational aspect, and is constructed \textit{post hoc} as the heuristic that indicates the insertion with the best performance among the results of $\beta$ heuristic.


\subsubsection{$\gamma$ Heuristic}\label{Metodo-gamma}

The $\gamma$ heuristic inserts a single edge between non-adjacent vertices with the minimum \textit{degree}. Its theoretical motivation is to reduce the average path length of $G$, which is a common measure of the efficiency of a network\cite{barabasi_2002, holme_2002, jiang_2011, ji_2016}. Given that the average path length is negatively correlated with connectivity, it is reasonable to suppose whether a heuristic that minimizes the average path length may also increase vertex reliability. As a derivative heuristic of $\gamma$, we postulate $\Gamma$ heuristic. This heuristic is ideal, and is constructed \textit{post hoc} as the heuristic that indicates the insertion with the best performance among the results of $\gamma$ heuristic.


\subsubsection{$\delta$ Heuristic}\label{Metodo-delta}

The $\delta$ heuristic inserts a single edge between non-adjacent vertices, one of which has the maximum \textit{degree} and the other with whom the first has the greatest geodesic distance. Its theoretical motivation is to reduce the diameter of the graph. The diameter is an inverse measure of the efficiency of a network, and it is related with the probability of a graph getting disconnected after the targeted removal of some of its vertices, especially in cases of graphs with non-homogeneous degree distributions\cite{barabasi_2000}. Given the similarity of this measure and reliability, it is reasonable to question whether a heuristic that minimizes the diameter may also increase vertex reliability.


\subsubsection{$r$ Heuristic}\label{Metodo-r}

The $r$ heuristic inserts a single edge between non-adjacent vertices \textit{randomly}. In the literature, it is referred to as \textit{Random Choice}. It is generally used as a reference or basis of comparison for the other heuristics\cite{beygelzimer_2005}.



\section{Computational Experiments}\label{EC}


\subsection{Dataset}\label{EC-Dados}

The generation, storage and analysis of the graphs were developed in \textit{Julia} v$1.7.1$ with the following packages and their respective versions: \textit{Combinatorics.jl} v$1.0.2$; \textit{DataFrames.jl} v$1.2.2$; \textit{GraphIO.jl} v$0.6.0$; \textit{Graphs.jl} v$1.4.1$; \textit{Polynomials.jl} v$2.0.17$; and \textit{SortingAlgorithms.jl} v$1.0.1$. All of them are Free and Open Source Software. The source code for the computational experiments is available at \textit{GitHub}\cite{webrepo_2022}. The hardware used was an Intel i7-7700K CPU with 32GB of RAM running a GNU/Linux 64-bit operating system.

In order to ensure a robust evaluation of the heuristics, a dataset was created with random connected graphs without pendant vertices of order $n \in \{10,\ldots,20\}$. For each order $n$, $2000$ graphs were generated: $1000$ Erd\H{o}s-Rényi (ER) graphs, $500$ Barabási-Albert (BA) graphs, and $500$ Watts-Strogatz (WS) graphs. In total, the dataset has $22000$ random graphs. Regardless of the model used to generate the random graph, each graph was tested to verify that it met the topological requirements and ensure no isomorphic graphs in the dataset.


\subsection{Comparison Criterion}\label{EC-criterio}

Consider $\mathcal{H}= \{ h_{1}, \ldots, h_{c} \}$, a set of $c$ heuristics, $\mathcal{G} = \{ G_{1}, \ldots, G_{N} \}$, a set of $N$ connected graphs, and $1 \leq k \leq c$.

The relative deviation index (RDI) is used to evaluate the performance of the heuristics\cite{kim_1993, moro_2017}. The definite integral of the vertex reliability polynomial $R_{N}(G, p)$ in the interval $[0,1]$ is selected as its score function, denoted by $F$. Thus, for each supergraph $Y_{ij}$ produced by an edge insertion of the $k$-th heuristic, there is an associated score value $F(Y_{ij})$. The higher the score, the more reliable a supergraph is when $p \in [0,1]$.



However, as some heuristics can produce more than one insertion of edge as a solution, the score is not exactly from the heuristic, but of an insertion and its associated supergraph. For each $G$, we define $F_{B}$ and $F_{W}$, respectively, as the maximum and minimum scores obtained by the supergraphs produced from the edge insertions of all the heuristics in $\mathcal{H}$. Let the $t$-th insertion of the $k$-th heuristic be denoted by $h_{k}^{(t)}$, where $l_{k}$ is the number of insertions produced by $h_{k}$ such that $1 \leq t \leq l_{k}$. Thus, the RDI of $h_{k}^{(t)}$ in $G$ is given by

    \begin{equation}
        \label{eq_rdij}
        RDI_{G, h_{k}^{(t)}} = \left\{ \begin{array}{cl}
            \frac{F_{B} - F(G + h_{k}^{(t)})}{F_{B} - F_{W}}, & \text{if $F_{B} - F_{W} \neq 0$} \\
            0, & \text{if $F_{B} - F_{W} = 0$}
        \end{array} \right.,
    \end{equation}

\noindent and the RDI of $h_{k}$ in $G$ is calculated as the arithmetic mean of the RDI of its associated edge insertions:
    
    \begin{equation}
        \label{eq_rdi}
        RDI_{G, h_{k}} = \frac{\sum^{l_{k}}_{t=1} RDI_{G, h_{k}^{(t)}}}{l_{k}}.
    \end{equation}

The mean relative deviation index (MRDI) compares heuristics between graphs in $\mathcal{G}$. The MRDI of $h_{k}$ is defined as

\begin{equation*}
    \label{eq_MRDI}
    MRDI_{\mathcal{G}, h_{k}} = \frac{\sum^{N}_{s=1} RDI_{G_{s},h_{k} }}{N},
\end{equation*}

\noindent where the heuristic with minimum MRDI is, on average, closer to the highest scores.


\subsection{Procedures}\label{EC-procedimento}

Therefore, to evaluate the heuristics, the following steps are considered:

\begin{enumerate}
    \item\label[step]{item:p01} Let $\mathcal{G}$ and $\mathcal{H}$, sets of $N$ connected graphs and $c$ heuristics. Then, for each $G \in \mathcal{G}$:
    
    \begin{enumerate}
        \item\label[step]{item:p01_01} Apply $h_{k}$ on $G$ and tag as $h_{k}^{(1)}, \ldots, h_{k}^{(l_{k})}$ the $l_{k}$ edge insertions produced by $h_{k}$, for $1 \leq k \leq c$.
    
        \item\label[step]{item:p01_02} Store all supergraphs $G + h_{k}^{(t)}$, for $1 \leq k \leq c$ and $1 \leq t \leq l_{k}$, produced by all the heuristics in $\mathcal{H}$.

        \item\label[step]{item:p01_03} Enumerate all the connected induced subgraphs of $G, G + h_{1}^{(1)}, \ldots, G + h_{c}^{(l_{c})}$.
    
        \item\label[step]{item:p01_04} Calculate the vertex reliability polynomial $R_{N}(G,p), R_{N}(G + h_{1}^{(1)}, p), \ldots, R_{N}(G + h_{c}^{(l_{c})}, p)$ as a function of $p$.
        
        \item\label[step]{item:p01_05} Calculate the RDI for each heuristic in $\mathcal{H}$ for $G$.
    \end{enumerate}
    
    \item\label[step]{item:p02} Calculate the MRDI for each heuristic in $\mathcal{H}$ for $\mathcal{G}$ and set as \textit{best} the heuristic with the lowest MRDI.
    
    \end{enumerate}

The power set of $V$, $\mathcal{P}(V)$, lists all possible $2^{n}-1$ induced subgraphs of $G$. Let $S \neq \emptyset$ be an element of $\mathcal{P}(V)$. Given that $G$ is connected, then any supergraph $Y_{ij}$ of $G$ is necessarily connected. Therefore, for any $S$, a connected subgraph induced by $S$ in $G$ is also connected in $Y_{ij}$, regardless of the edge insertion. Thus, none of these subgraphs need to be tested more than once. The number of connected induced subgraphs can only increase for the supergraphs of $G$. If the subgraph induced by $S$ is disconnected in $G$ and $\{ v_{i}, v_{j} \} \not\subset S$, then it is necessarily disconnected in $Y_{ij}$. That is, an edge insertion between vertices that are not in $S$ does not change the connectivity of the induced subgraph. Therefore, it is only needed to test the disconnected induced subgraphs that contain both $v_{i}$ and $v_{j}$ for each $Y_{ij}$. Once all disconnected induced subgraphs of $G$ that become connected induced subgraphs in $Y_{ij}$ are counted, we obtain the vertex reliability polynomial for $Y_{ij}$.

The RDI of each basic heuristic in $\mathcal{H}$ for each $G$ is known at the end of the procedure. With this data, one can construct the derivative heuristics, calculate the MRDI, and then analyse the performance of all the heuristics. 

For the derivative heuristics, we applied an approximate one-sided Wilcoxon signed-rank test with Bonferroni correction for multiple comparisons and point-biserial correlation coefficient $r$ for the effect size. The null hypothesis is that the median of pairwise differences is less than or equal to zero; the alternative hypothesis is that the median of pairwise differences is greater than zero. It is possible to identify whether the distributions of the scores between heuristics are significantly distinct and if there is any stochastically dominant heuristic through these tests\cite{wilcoxon_1992}. Two of the derivative heuristics are \textit{post hoc}, so their comparison with $\Phi$ is quite conservative, favoring the null hypothesis.


\subsection{Results}\label{EC-Analise}

For conciseness, the analysis presented here is limited to the case in which all graphs are aggregated by their model -- ER, BA, or WS -- and order. The analysis of the disaggregated data exhibited similar results in terms of performance.

The quality of a heuristic depends not only on its performance but also on its computational cost. \cref{tabela-benchmark} presents an empirical evaluation of the computational time (in milliseconds) of heuristics in $\mathcal{H}$ for two sets of graphs with orders $15$ and $20$, each with $1000$ ER random graphs.

\begin{table}[h]
    \caption{Empirical computational time for the heuristics $[ms]$}
    \label{tabela-benchmark}
    \centering
    \scalebox{0.8}{
        \begin{tabular}{c c l l l l}
            \hline
            \hline

            $\vert V \vert$ & Heuristic & Min & Max & Median & Mean $\pm$ SD\\
            \hline
            
            15 & $\alpha$ & $0.601$ & $2.809$ & $0.885$ & $0.931 \pm 0.291$\\
            
            & $\beta$ & $0.063$ & $3.434$ & $0.077$ & $0.079\pm 0.001$\\
            
            & $\gamma$ & $0.015$ & $0.047$ & $0.019$ & $0.019 \pm 0.001$\\
            
            & $\delta$ & $0.033$ & $3.923$ & $0.070$ & $0.104\pm 0.226$\\
            
            & $\varphi$ & $0.086$ & $2.816$ & $0.107$ & $0.110 \pm 0.087$\\
            
            & $r$ & $0.010$ & $0.020$ & $0.012$ & $0.012\pm 0.001$\\
            
            & $\Phi$ & $0.104$ & $2.778$ & $0.125$ & $0.133\pm 0.120$\\
            
            20 & $\alpha$ & $1.726$ & $4.770$ & $2.274$ & $2.398 \pm 0.509$\\

            & $\beta$ & $0.111$ & $3.397$ & $0.120$ & $0.137 \pm 0.176$\\

            & $\gamma$ & $0.027$ & $0.220$ & $0.033$ & $0.034 \pm 0.006$\\
            
            & $\delta$ & $0.044$ & $2.346$ & $0.056$ & $0.081 \pm 0.143$\\
            
            & $\varphi$ & $0.142$ & $2.439$ & $0.175$ & $0.189 \pm 0.123$\\
            
            & $r$ & $0.018$ & $0.032$ & $0.024$ & $0.024 \pm 0.001$\\
            
            & $\Phi$ & $0.173$ & $2.556$ & $0.201$ & $0.218 \pm 0.146$\\
            
            \hline
            \hline
        \end{tabular}}
\end{table}

\cref{tabela-desempenho} shows the following summary measures for each heuristic: \textbf{(a)} \textit{Insertions} is the total number of insertions produced by the heuristic; \textbf{(b)} \textit{Best (Single)} is a pair of values, where the first is the number of insertions produced by the heuristics that have reached the highest vertex reliability value among all insertions indicated by the set $\mathcal{H}$ for the same graph, and the second (in parentheses) is the number of graphs in which the heuristic produced at least one insertion that obtained one of these higher values; \textbf{(c)} \textit{MRDI} is the average relative deviation index of the heuristic; and \textbf{(d)} \textit{SD RDI} is the standard deviation of the RDI of the insertions produced by the heuristic. In addition, cells are coloured using a sequence of colours ordered from red to blue (\crule[Pos09]{1.2ex}{1.2ex}\crule[Pos08]{1.2ex}{1.2ex}\crule[Pos07]{1.2ex}{1.2ex}\crule[Pos06]{1.2ex}{1.2ex}\crule[Pos05]{1.2ex}{1.2ex}\crule[Pos04]{1.2ex}{1.2ex}\crule[Pos03]{1.2ex}{1.2ex}\crule[Pos02]{1.2ex}{1.2ex}\crule[Pos01]{1.2ex}{1.2ex}), where the values correspond to a rank classification for each column. A heuristic with the best performance has most of its values in the darkest blue (\crule[Pos01]{1.2ex}{1.2ex}), especially those related to MRDI.

\begin{table}[h]
    \caption{Performance of the edge-insertion heuristics}
    \label{tabela-desempenho}
    \centering
        \scalebox{0.8}{
        \begin{tabular}{c c c c c}
            \hline
            \hline
            Heuristic & Insertions & Best (Unique) & MRDI & SD RDI\\
            \hline
        
            $\alpha$ &
            \cellcolor{Pos05!60} $23615$ &
            \cellcolor{Pos05!60} $11959 \ (11325)$ &
            \cellcolor{Pos04!60} $0.115049$ &
            \cellcolor{Pos03!60} $0.194493$\\

            $\beta$ &
            \cellcolor{Pos07!60} $53835$ &
            \cellcolor{Pos06!60} $6569 \ (6216)$ &
            \cellcolor{Pos07!60} $0.515709$ &
            \cellcolor{Pos09!60} $0.359231$\\

            $\gamma$ &
            \cellcolor{Pos08!60} $74436$ &
            \cellcolor{Pos03!60} $14062 \ (13447)$ &
            \cellcolor{Pos06!60} $0.315827$ &
            \cellcolor{Pos05!60} $0.287909$\\

            $\delta$ &
            \cellcolor{Pos09!60} $165483$ &
            \cellcolor{Pos08!60} $5548 \ (5198)$ &
            \cellcolor{Pos08!60} $0.649696$ &
            \cellcolor{Pos07!60} $0.297809$\\

            $\varphi$ &
            \cellcolor{Pos06!60} $23995$ &
            \cellcolor{Pos01!60} $14870 \ (14209)$ &
            \cellcolor{Pos02!60} $0.052739$ &
            \cellcolor{Pos02!60} $0.105383$\\

            $r$ &
            \cellcolor{Pos01!60} $22000$ &
            \cellcolor{Pos09!60} $743 \ (743)$ &
            \cellcolor{Pos09!60} $0.678662$ &
            \cellcolor{Pos08!60} $0.309818$\\

            $B$ &
            \cellcolor{Pos01!60} $22000$ &
            \cellcolor{Pos06!60} $6216 \ (6216)$ &
            \cellcolor{Pos05!60} $0.260597$ &
            \cellcolor{Pos06!60} $0.290210$\\

            $\Gamma$ &
            \cellcolor{Pos01!60} $22000$ &
            \cellcolor{Pos03!60} $13447 \ (13447)$ &
            \cellcolor{Pos03!60} $0.093376$ &
            \cellcolor{Pos04!60} $0.194638$\\

            $\Phi$ &
            \cellcolor{Pos01!60} $22000$ &
            \cellcolor{Pos02!60} $14187 \ (14187)$ &
            \cellcolor{Pos01!60} $0.042533$ &
            \cellcolor{Pos01!60} $0.088694$\\
        
            \hline
            \hline
        \end{tabular}}
\end{table}

\begin{figure}[ht]
    \centering
    \includegraphics[width=0.35\textwidth]{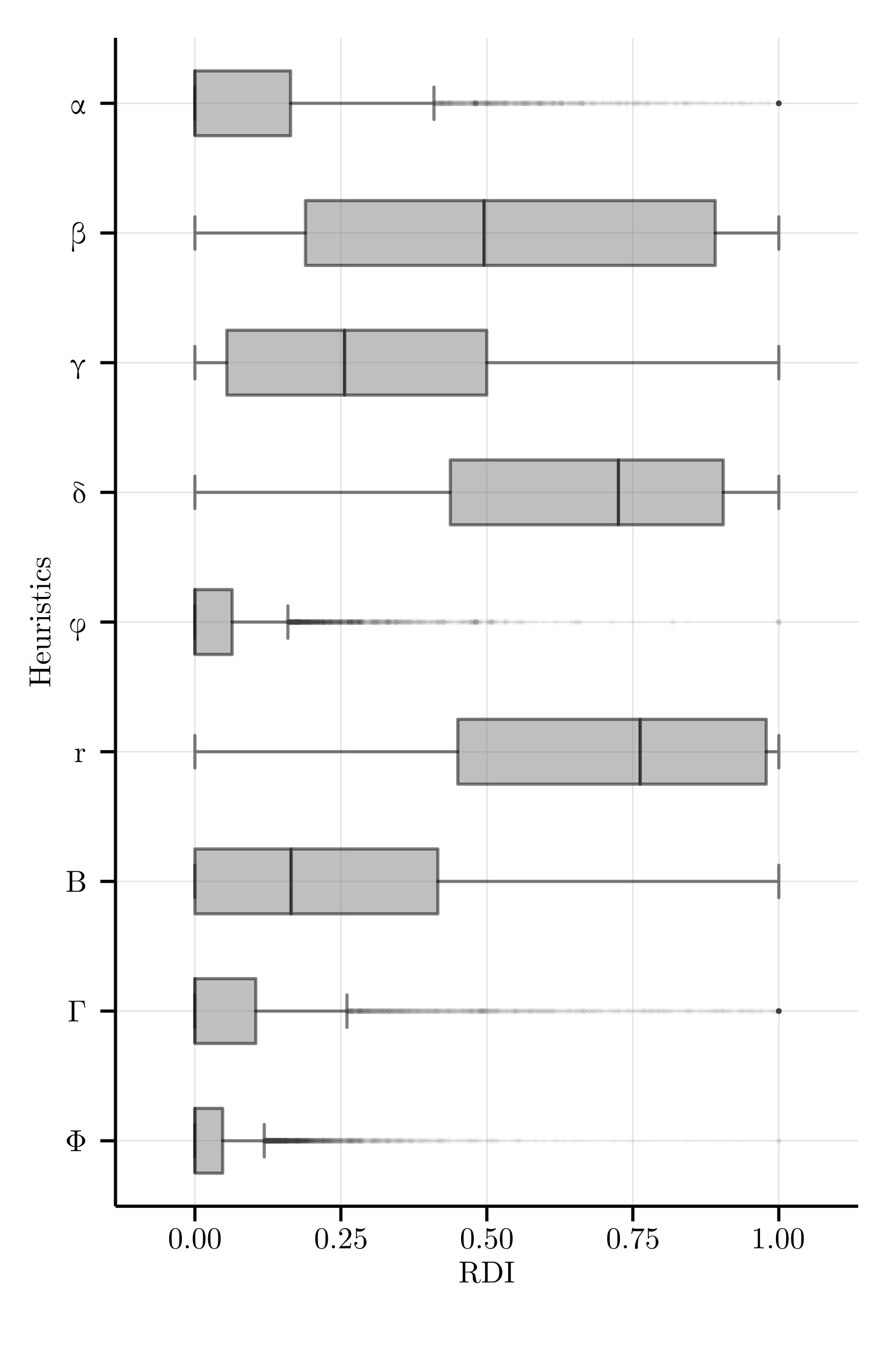}
    \caption{Boxplot of the edge-insertion heuristics' RDI}
    \label{fig-boxplot}
\end{figure}

In addition, the distribution of the RDI of the insertions produced by each heuristic can be seen in \cref{fig-boxplot}. Some elements of the boxplots are not visible in some heuristics because they are flattened around $0$.

The $\varphi$ heuristic presented the best performance regarding MRDI among all basic heuristics in the dataset of graphs. It also has the lowest standard deviation of the RDI, which is a desirable property. Accordingly, $\varphi$ is comparatively better than the other basic heuristics tested by the criterion we have established.

In the case of the derivative heuristics, it can be seen that $B$ and $\Gamma$ present lower values of MRDI than their respective basic heuristics, as expected. In turn, $\Phi$ exhibits a modest improvement over MRDI and SD RDI. Nevertheless, the number of graphs in which $\Phi$ produced the best result had a slight reduction compared to its basic heuristic. The results of the Wilcoxon signed-rank tests for each pair of derivative heuristics are shown in \cref{tabela-wilcoxon} with the following columns: \textbf{(a)} $r$ is the effect size given by the point-biserial correlation coefficient; \textbf{(b)} $T^{*}$ is the standardized test statistic; and \textbf{(c)} $p$-value is the probability of a value as extreme as the observed given the null hypothesis. These results provide sufficient evidence for $\Phi$ to be ranked as the best heuristic among the derivative heuristics.

\begin{table}[!t]
	\caption{(Approximated) Wilcoxon signed-rank tests for the heuristics}
	\label{tabela-wilcoxon}
	\centering
	 \scalebox{0.8}{
    \begin{tabular}{m{2cm} m{1cm} m{1cm} m{2cm}}
	    \hline
	    \hline
	    Heuristics & $r$ & $T^{*}$ & $p$-value\\
	    \hline
	    ($\Phi$, $B$) &	$0.9618$ & $94.3794$ & $\ll 0.0001$\\
	    
	    ($\Phi$, $\Gamma$) & $0.6760$ & $28.4584$ & $\ll 0.0001$\\
	    
	    ($\Gamma$, $B$) & $0.8855$ & $76.1101$ & $\ll 0.0001$\\
	    
	    \hline
	    \hline
    \end{tabular}}
\end{table}



\section{Conclusion}\label{Conclusao}

In this paper, we proposed an adaptation of a spectral heuristic for the problem of vertex reliability of a single edge insertion. We named the basic adaptation as $\varphi$ and its refinement as $\Phi$. Both were compared in a set of random graphs with other heuristics, some from the literature, while others are purely idealized.

Following the defined comparison criterion, the heuristics proposed have had been the best performance. The heuristic $\Phi$ has a higher computational cost than its basic form. However, its most costly computation is the tiebreak criterion. In practice, the computational time is expected to be, on average, very close. Therefore, it is reasonable to conclude that $\Phi$ provides a good enough answer to the research problem presented here and should be used in instances where it is feasible.

For future works, the cases where $\varphi$ outperforms $\Phi$ could be analysed. If this behavior is related to some topological properties of the graphs, then $\Phi$ can be further improved with minor adjustments. There are also good reasons to believe that it is possible to improve the performance of $\Phi$ by identifying other tiebreak criteria for the multiplicity of edge insertions. . Alternatives may involve, for example, calculating the Fiedler vectors of the supergraphs resulting from the insertions of edges to identify with more accurately, via Fiedler's Theorem, the supergraph with the largest number of connected induced subgraphs. Considering the theoretical motivation developed here for the $\varphi$ heuristic, is desirable to explore more properties of the Fiedler vector in the context of reliability in general.







\bibliographystyle{ieeetr}
\bibliography{References}

\ifCLASSOPTIONcaptionsoff
  \newpage
\fi

%


\vspace{-1.0cm}

\begin{IEEEbiography}[{\includegraphics[width=1in,height=1.25in,clip,keepaspectratio]{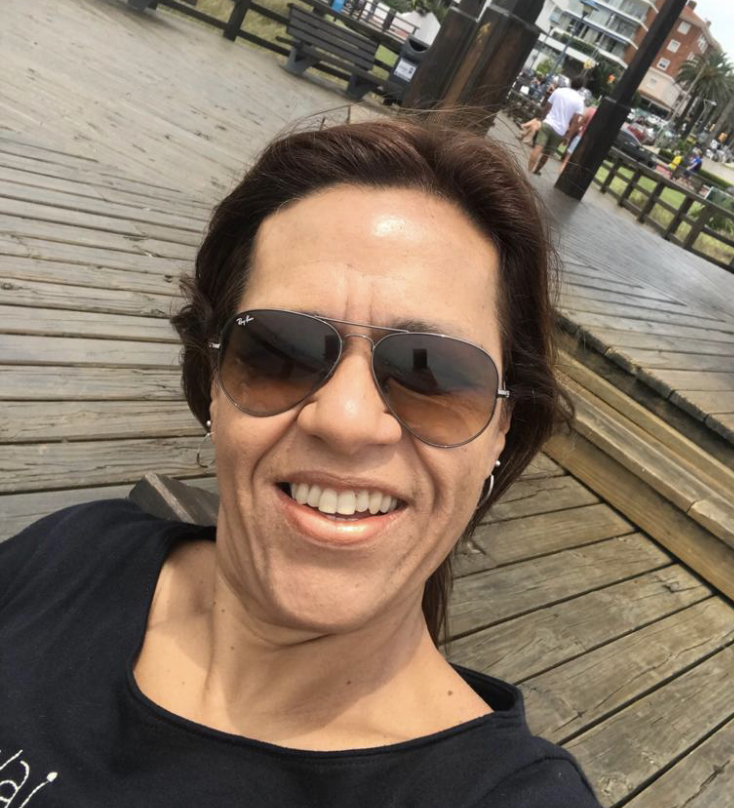}}]{Carla Silva Oliveira}
    B.Sci. and M.Sci. in Mathematics from the Fluminense Federal University (UFF) and Ph.D. in Production Engineering from the Federal University of Rio de Janeiro (UFRJ). She is a researcher at the National School of Statistical Sciences (ENCE/IBGE) and has experience in Graph Theory and Spectral Graph Theory.
\end{IEEEbiography}

\vspace{-1.0cm}

\begin{IEEEbiography}[{\includegraphics[width=1in,height=1.25in,clip,keepaspectratio]{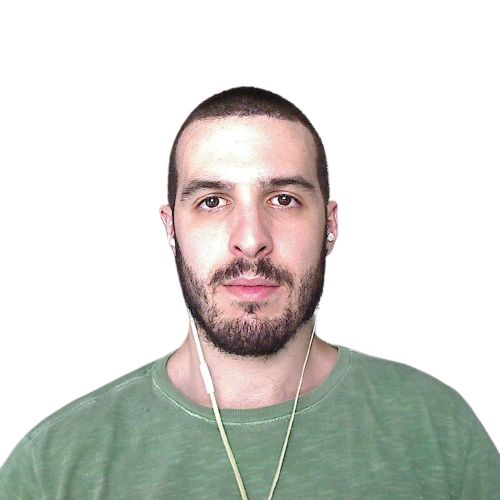}}]{Fausto Marques Pinheiro Junior}
    B.A. in Law from the Federal University of Rio de Janeiro (UFRJ), B.Sci. in Statistics from the National School of Statistical Sciences (ENCE/IBGE), M.A. in Theory and Philosophy of Law from the Rio de Janeiro State University (UERJ) and is currently pursuing a M.Sci. in Applied Mathematics at the State University of Campinas (Unicamp).
\end{IEEEbiography}

\vspace{-1.0cm}

\begin{IEEEbiography}[{\includegraphics[width=1in,height=1.25in,clip,keepaspectratio]{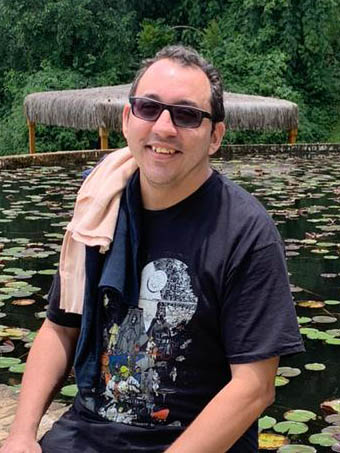}}]{José André de Moura Brito}
    B.Sci. in Mathematics from the Federal University of Rio de Janeiro (UFRJ), M.Sci. (1999) and Ph.D. (2004) in Systems and Computer Engineering (Optimization) from COPPE/UFRJ. He was a postdoctoral research fellow in Optimization at the Fluminense Federal University (UFF) and is currently a professor at the National School of Statistical Sciences (ENCE/IBGE), where he teaches undergraduate courses. He has experience in the areas of Optimization, Statistics and Computing.
\end{IEEEbiography}


\end{document}